\documentclass[aps,pre]{revtex4}
\usepackage{graphicx}
\usepackage{hyperref}

\begin{document}

\title{Adaptive Dynamic Congestion Avoidance with Master Equation} 


\author{Mehmet S{\"u}zen}
\email[]{suzen@fias.uni-frankfurt.de, mehmet.suzen@physics.org}
\affiliation{ Frankfurt International Graduate School for Science, J.W. Goethe University, Ruth-Moufang-Str. 1 D-60438 Frankfurt am Main, Federal Republic of Germany}

\author{Ziya S{\"u}zen}
\email[]{zsuzen@acm.org, ziya@suzen.net}
\affiliation{Member IEEE/ACM, 99 Landcliffe Drive, Heelands, Milton Keynes, MK13 7LB, United Kingdom}

\begin{abstract}
This paper proposes an adaptive variant of Random Early Detection (RED) gateway 
queue management for packet-switched networks via a discrete state analog of the non-stationary Master Equation
i.e. Markov process.  The computation of average queue size, which appeared in the original RED algorithm, is
altered by introducing a probability $P(l,t)$, which defines the probability of having $l$ number
of packets in the queue at the given time $t$, and depends upon the previous state of the queue. This
brings the advantage of eliminating a free parameter: queue weight, completely. Computation of transition rates and
probabilities are carried out on the fly, and determined by the algorithm automatically.  Simulations with unstructured packets illustrate the method, the performance of the adaptive variant of RED algorithm, and the comparison with the standard RED.
\end{abstract}


\date{August 2008}

\maketitle
\section{Introduction}
Dynamics in the self-driven, many particle systems have been related to traffic flow problems,
and attracted the interest of many different disciplines \cite{helbing01a}. Traffic breakdown in these systems is a prominent example. 
These studies mostly focused on jams of freeway traffic. Similar approaches have been taken along these lines for congestion avoidance in computer networks, especially congestion control on the Internet.  Techniques of mathematical and statistical physics are suitable and applicable, 
since these problems have universal dynamics \cite{mukherjee91a,srikanthan2004mic}.

Congestion avoidance, or control, aims at preventing the dynamic breakdown of the traffic and 
insuring a fair sharing of resources. For the case of packet-switching networks, bottlenecks 
may frequently occur in gateways. Previous studies, that addressed congestion avoidance in 
gateways, have focused on mechanisms which keep throughput high and average queue size low.
Some of these mechanisms are designed to work with gateways or coupled to gateway scheduling algorithms \cite{jacobson95a}. 

The most notable and widely used algorithm is Random Early Detection (RED) gateways \cite{floyd93a}. 
The primary mechanism that RED provides, allows queue size growth as long as a statistical probability measure is
satisfied, otherwise incoming packets are dropped. In this way the gateway adjusts queue buffer space for fair 
traffic share. Dynamics of congestion control models and RED gateways were rigorously investigated under different settings
in the large literature on the subject, We follow a different path and present a direct
adaptive variant to the original RED algorithm with new interpretation of the state of the queue, which
is characterised by its length, is coupled with a Master Equation.  

\section{A Master Equation RED}
  The investigation of many physical time-dependent processes starts with the statistical description
  based on well studied generalised Master Equation \cite{vankampen1992spc}.

   The discrete state analog of the Master Equation as 
   a differential difference equation in one-dimension is 
   \begin{eqnarray}
    d P(l,t) / dt  & = & a_{l,l+1} P(l+1,t) + a_{l,l-1} P(l-1,t)-  \nonumber  \\
                   &   & (a_{l-1,l} + a_{l+1,l}) P(l,t)
   \end{eqnarray}
   where $P(l+j,t)$ is the probability that the system is in the state $(l+j)$ at time $t$
   and where $a_{l,j}$ are the transition rates from state $j$ to $l$. Addressing the theory behind Master Equation
   goes beyond the scope of the present study.
   
   In the original RED algorithm, the computation of instantaneous average queue size, $avg$,
   is determined dynamically by an exponential weighted moving average when the queue is non-empty

    $$ avg = (1- w_{q}) avg + w_{q} q$$

   where $w_{q}$ is the queue weight, a free parameter, and $q$ is the current queue size .
   
   The primary method in the Master Equation Random Early Detection (mRED) algorithm is interpreting
   probability evolution $P(l,t)$ as the probability of having $l$ number of packets in the queue
   at given time $t$. This probability comes into play in finding the average
   queue size, $avg$, where $q$ is the current queue size at time $t$
   \begin{equation}
     avg=q P(q,t) 
   \end{equation}
   Next, $a_{l,j}$ is the transition rate from state $j$ to $l$, which determined by using  
   free energy \cite{vankampen1992spc} of a given state.  Here, the free energy measure is 
   taken to be an absolute of sum of aimed queue length and average queue length in the previous
   state. Hence, forward or backward transition rates are computed as follows
    \begin{equation}
      a_{j,l}=exp(-|j+avg|)
    \end{equation}
   since we are in Markov regime, $j$ can only take values $l+1$ and $l-1$, for forward and 
   backward rates respectively. The average queue size plays an important role here, namely the information
   from the previous state of the queue.  This value changes at each time step, so one needs to 
   update all probabilities and the transition rates accordingly. The exponential in the transition rates 
   is a reminiscent of the Boltzmann factor of Statistical Physics. 

   As an illustrative example. If 6 and 4 were forward and backward queue length, 5 is the current queue size, 
   and if $avg$ is 0.3, then simply  symmetric transition rates will be a(5,6)=0.0049 and a(6,5)=0.0018.
   However, the following ratio between symmetric index rate constants is valid at the {\it weak detailed balance}
   condition (when system is at near equilibrium)

    \begin{equation}
     a(j,j-1) / a(j-1,j) = exp (-\Delta E )
      \label{diff}
    \end{equation}
 
   $\Delta E$ is the energy difference between states $j$ and $j-1$.  Since, the measure of energy in queue system 
   is expressed with the $avg$ and the current queue length. Ideally $\Delta E$ should be zero, if current queue 
   length $j$ or its neighbors near it are in the optimal value, analogy with a minimum. The energies in a given
   transition $j \to j-1$ or $j-1 \to j$ expressed as exponential of the starting state plus the
   average queue length, $exp(-|j+avg])$ and $exp(-|j-1+avg|)$, respectively.  And, the only condition 
   to make this ratio 1, is when $j-1$ or $j$ approaches the value of average queue size. So actually, 
   imposing the condition in the eq. \ref{diff} is to drive the current queue length to average queue length. 

   The {\it weak detailed balance} condition, that $\Delta E$ should be zero, impose an other condition,
 
    \begin{equation}
    a(j,j-1) / a(j-1,j) =  a(j,j+1) / a(j+1,j)
      \label{diff2}
    \end{equation}

   So imposing this condition implies we can take as the following rates equal, $a(j,j-1) =  a(j,j+1)$. Important
   point is we do not take eqns \ref{diff} and \ref{diff2} directly in solving master equations rather, we would
   like to drive the queue behaviour into this conditions. Moreover, this type of  formulations is 
   intrinsically phenomenological and constructed with making analogy with the queue behaviour 
   as an approximation.  
   
   Implementation of mRED must follow the same path as RED with the different computation
   strategy for the average queue size. This can be achieved by solving the differential 
   difference Master Equation numerically on every packet arrival or on specified 
   time intervals. The frequency of updating transition rates and probabilities
   can be decided. In our implementation, we solved the equation on every packet
   arrival with the information of current queue size. The general algorithm for mRED is given 

\newpage
\begin{verbatim}
for each packet arrival
  Update the set of transition rates using 
   previous average queue size
  Solve numerically the set of Master Equations 
   for all probability values P(l,t)
  Calculate the average queue size by using P(q,t)
  if minth <= avg < maxth
       Calculate marking probability 
       with marking probability :
          mark the arriving packet
    else if maxth <= avg
          mark the arriving packet 
\end{verbatim}

   If the buffer size is $N$, including the zero size queue, then one needs to store, 
   $2N-2$ number of transition rates and $N$ number of probability values (because one needs to know forward and backward transition
   rates for $N$ state, zero to buffer size, but for queue size zero there is no backward transition rate and 
   for queue size $N$ there is no forward transition rate, so mRED needs to store and update $2N-2$ transition rates).  
   All values can be initialised randomly, and the procedure updates the values in response to previous 
   behaviour of the queue.
   
   Let us remember that the current queue size is $q$. The probability of having $q$ packets in the queue
   at the current moment is extracted as follows. First, a number of $N$ probabilities : 
   $P(0,t)$,$P(1,t)$,...,$P(N,t)$ are computed solving the Master equation. Then, knowing the current 
   queue size $q$, we extract $P(q,t)$ from the list.

\section{Simulations}

  Simulations were carried out with a simulation package, that we have developed \cite{mred07}. Two different network 
  settings have been simulated, where some of the original parameters reported in \cite{floyd93a} were used. Those are threshold values 
  $minth=5$ and $maxth=15$, for minimum and maximum queue threshold respectively and $maxp=1/50$, the threshold 
  for marking probability (marked packets are dropped by the gateway).  We first studied a network for 
  which the highest possible packet arrival is slightly higher than the maximum queue threshold ($maxth$). Next, we 
  investigated a network with bursty traffic where the highest packet arrival rate can be as high as $3$ times 
  the maximum queue threshold.
  

  The initial step in setting up the network is assigning a number of hosts that generated  the 
  packets to the gateway. At each time step, randomly selected hosts were sending packets
  that were processed by the queue management system. If the packet is marked for drop, the corresponding host gets the drop contribution.
  The Master Equation routine is invoked whenever the average queue size, $avg$,  was needed and the current queue size $q$,
  was provided to the algorithm. First, transition rates are computed for all the possible queue sizes,
  from zero size to buffer size, following the same notation (for $a_{l,j}$ values, indexes correspond to
  each other in the following way, $j=1,..N-1$ and $l=2,..,N$ for forward transitions, and $j=2,..N$ and $l=1,..,N-1$ for 
  backward transitions, respectively). Afterwards, all possible queue size probabilities are computed by 
  solving a Master Equation numerically (Forward Euler Method) for a later time with small enough $\Delta t$
    \begin{eqnarray}
     P(l,t+\Delta t) & = &  ( a_{l,l+1} P(l+1,t) + a_{l,l-1} P(l-1,t)-  \nonumber \\
                     &   & (a_{l-1,l} + a_{l+1,l}) P(l,t))  \Delta t + P(l,t)  
    \end{eqnarray}
  Hence, the probability of having current queue size $q$ is returned back to the main mRED routine to compute the 
  current average queue size.

  Here, the buffer size was big enough to prevent any drops due to buffer overflow. The packet
  drop rates do not depend upon the buffer size.  But, buffer size plays an important role in 
  updating probabilities in mRED. It determines the propagation of probability values. Remember that
  any probability value $P(l,t)$ also depends upon $P(l-1,t)$  and $P(l+1,t)$. In this sense, the number of 
  possible different queue length must be large enough to have a better statistics for probability values. 
  For that reason we choose the value $N=500$ as a test case which is not too small. All the results reported used
  this value for the buffer size $N$.

  We have used a time step of $\Delta t=0.01$ which gives sufficiently accurate results for testing
  purposes. We have used 6 hosts connected to a gateway in two different network traffic conditions. Network 1 and 2 have 
  assigned the following packet rates per unit time for hosts, ($2,3,2,1,6,4$) and ($14,7,4,5,12,3$), 
  and the maximum possible rate of $18$ and $45$, respectively. 



\section{Discussion}

      \begin{figure}
       \centering
        \includegraphics[angle=0,width=0.5\textwidth]{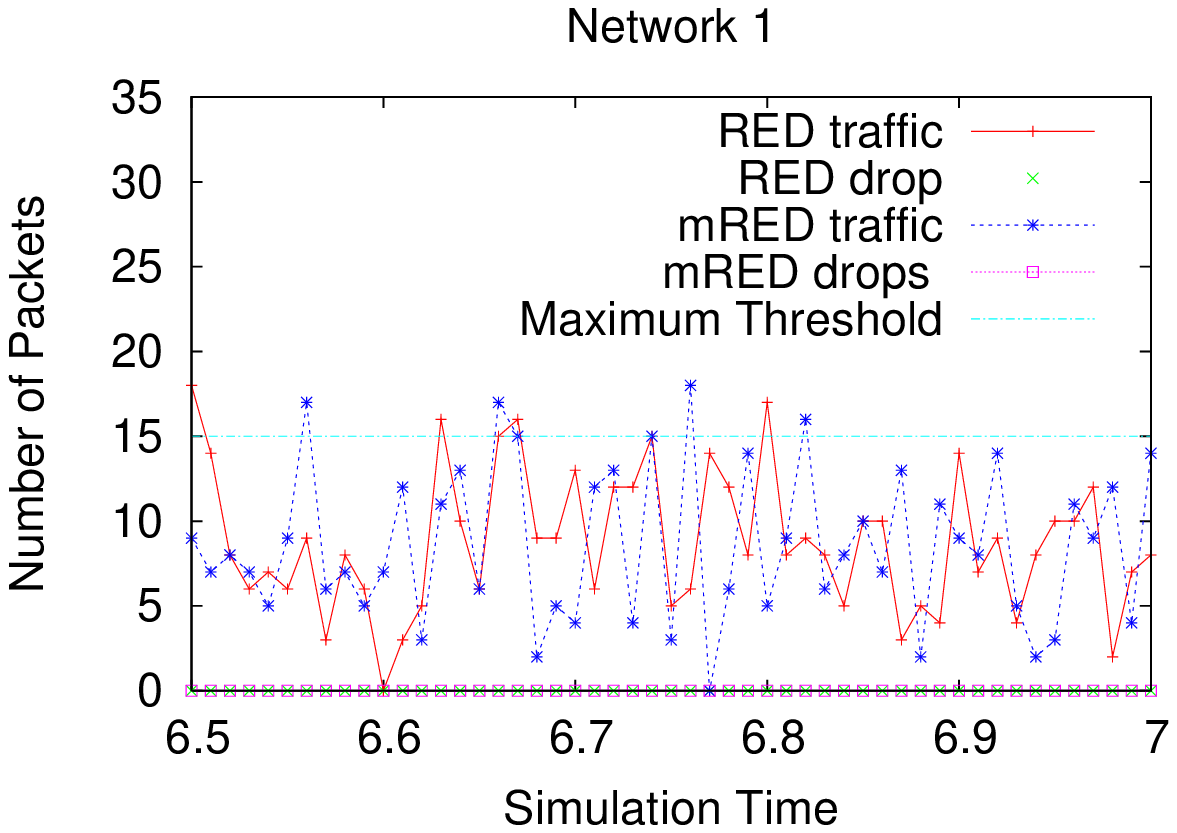}
        \includegraphics[angle=0,width=0.5\textwidth]{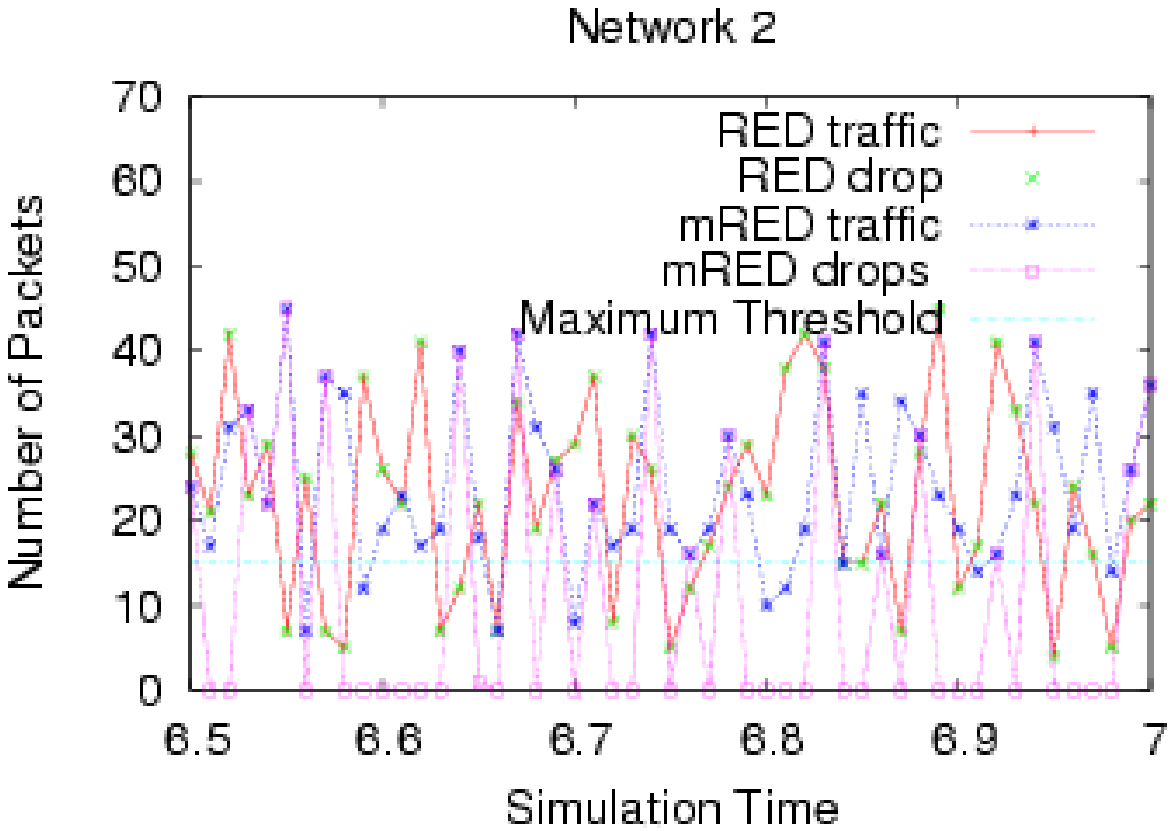}
         \caption{The total traffic and packet drop for both sample Network 1 and 2. 
		  The relative traffic utilisation drawn from this data, one minus 
		  the ratio of total drop to total traffic.  Simulation carried on 10000 steps. 
		  Here only a portion is shown. Network 2 mimics heavily congested traffic
		  situation. }
         \label{traffix}
      \end{figure}

The comparison of total traffic and drops for every step are reported in Figure \ref{traffix}. 
We can see that behaviour of both mRED and RED is quite similar. This is an important point,
because it reflects and hints that mRED do not introduce any artifact on the original design of RED.
The traffic performance of mRED must be compared with the proposed adaptive RED  \cite{floyd2001ara},  
which removes other two threshold parameters. Moreover mRED can be combined with adaptive RED 
\cite{floyd2001ara}, which such a scheme will remove all the free parameters. These points must 
be address in a future study.

Introduction of an arbitrary discritization time step changes the smoothing time-scale. 
This can be overcome by choosing smaller time intervals for the Master Equation solver,
such as orders of milliseconds. In practice millisecond is the conventional time-scale in packet-switched networks. 
Also, time interval is hidden in computation of transition rates in the formulation we have used, 
and it must be introduced for real-time implementations.

The computational performance of mRED algorithm depends on the buffer size, not to confuse with the 
traffic performance measured with relative traffic utilisation.  The implementations issues are
not the scope of the current paper but a simple analysis hints feasibility of implimentation. 
While number of algebraic operations for solving master equations grow linearly with the queue 
buffer size. For example a rough estimate of brute force implementation of mRED can be sketch as follows.
Assuming buffer size of 2000 which is 20 times larger then a typical router packet buffer, then,
the total operations needed for whole mRED is 43 986,  which is about $\sim 5$ KFLOPS.
This number of operation can be handled with modern processor like x86 in couple of milliseconds or 
less (assuming that processor can handle more then 2 MFLOPS). Moreover memory requirement is only 
about $\sim 100$ KB for whole procedure, and grows linearly with increasing buffer size. 

\section{Conclusion}

The primary advantage of using Master Equation is the elimination of the queue weight parameter $w_{q}$.
This is clearly a desirable and beneficial approach under circumstances 
where determination of $w_q$ is not straight forward.  However, the further investigation of mRED's
behaviour and its comparison with adaptive RED \cite{floyd2001ara} is needed, by using real 
packets and protocol environment, such as TCP and FTP.

\section*{Acknowledgement}

We are grateful to Dr. Raul C. Muresan and
anonoymous reviewers for critical reading of the manuscript.  
We thank Mehmet Alpturk in Nicosia, Cyprus for providing us an opportunity
to work on performance issues in border routers.

\bibliographystyle{unsrt}
\bibliography{research}

\begin{thebibliography}{1}

\bibitem{helbing01a}
D.~Helbing.
\newblock Traffic and related self-driven many-particle systems.
\newblock {\em Reviews of Modern Physics}, 73(4):1067--1141, 2001.

\bibitem{mukherjee91a}
A.~Mukherjee and J.C. Strikwerda.
\newblock Analysis of dynamic congestion control protocols: a fokker-planck
  approximation.
\newblock {\em ACM SIGCOMM Computer Communication Review}, 21(4):159--169,
  1991.

\bibitem{srikanthan2004mic}
R.~Srikanthan.
\newblock {\em The Mathematics of Internet Congestion Control}.
\newblock Birkh{\"a}user, 2004.

\bibitem{jacobson95a}
V.~Jacobson.
\newblock Congestion avoidance and control.
\newblock {\em ACM SIGCOMM Computer Communication Review}, 25(1):157--187,
  1995.

\bibitem{floyd93a}
S.~Floyd and V.~Jacobson.
\newblock Random early detection gateways for congestion avoidance.
\newblock {\em IEEE/ACM Transactions on Networking (TON)}, 1(4):397--413, 1993.

\bibitem{vankampen1992spc}
NG~van Kampen.
\newblock {\em Stochastic Processes in Chemistry and Physics}.
\newblock North-Holland, Amsterdam, 1992.

\bibitem{mred07}
M.~Suzen and Z.~Suzen.
\newblock Simple master equation red simulator: mred.
\newblock {\em GPLed Simulation code, available at
  \url{http://code.google.com/p/mred/}}, 2005.

\bibitem{floyd2001ara}
S.~Floyd, R.~Gummadi, S.~Shenker, et~al.
\newblock Adaptive red: An algorithm for increasing the robustness of red’s
  active queue management.
\newblock {\em Preprint, available at
  \url{http://www.icir.org/floyd/papers.html}}, 2001.

\end{thebibliography}
\end{document}